\documentclass[thmsb,notitlepage,11pt]{article}

\usepackage[ULforem,normalbf]{ulem}
\usepackage{graphicx}
\usepackage{amsfonts,amsmath,amsthm, amstext,amssymb}
\usepackage{enumerate}
\usepackage{pgf, tikz}
\usepackage[round,authoryear]{natbib}
\usepackage{epic}
\usepackage{color}
\usepackage{changepage}
\usepackage{multirow}
\usepackage{afterpage}
\usepackage{lscape}
\usepackage[font=small,labelfont=bf]{caption}
\usepackage{lineno}
\usepackage{cases}
\usepackage{array}
\usepackage[round,authoryear]{natbib}

\usepackage[english]{babel}
\usepackage{makeidx}
\usepackage[compatibility=false]{caption}
\usepackage{subcaption}
\usepackage{mathpazo} 
\usepackage{multimedia}

\renewcommand{\cite}[1]{\citep{#1}}

\def \R{\ensuremath{\mathbb{R}}}

\def \0{\ensuremath{\mathbf{0}}}
\def \1{\ensuremath{\mathbf{1}}}

\renewcommand{\bar}{\overline}

\def\bq{\begin{equation}}
\def\eq{\end{equation}}
\def\ba{\begin{eqnarray}}
\def\ea{\end{eqnarray}}
\def\bas{\begin{eqnarray*}}
\def\eas{\end{eqnarray*}}

\begin{document}

\title{\textbf{{\Large Models, Markets, and the Forecasting of Elections\thanks{We thank G. Elliott Morris for providing us with daily forecast data from the \textit{Economist} model, Parker Howell at \textit{PredictIt} for access to daily price data, and Pavel Atanasov, David Budescu, Jason Pipkin, David Rothschild, and partcipants at the 2021 DIMACS Workshop on Forecasting for comments on an earlier version. Rajiv Sethi thanks the Radcliffe Institute for Advanced Study at Harvard University for fellowship support.} \bigskip }%
}}

\author{Rajiv Sethi\thanks{%
Department of Economics, Barnard College, Columbia University, and the Santa
Fe Institute} \qquad
Julie Seager\thanks{Department of Economics, Barnard College, Columbia University} 
\qquad Emily Cai\thanks{Fu Foundation School of Engineering and Applied Science, Columbia University} 
\\ Daniel M. Benjamin\thanks{Information Sciences Institute, University of Southern California}
\qquad Fred Morstatter\thanks{Information Sciences Institute, University of Southern California}
}
\maketitle

\begin{abstract}
  \noindent We examine  probabilistic forecasts for battleground states in the 2020 US presidential election, using daily data from two sources over seven months: a statistical model based on election polls and fundamentals, and prices from a prediction market. We find systematic differences in accuracy over time with the market performing better months before the election, and the model performing better as the election approached. A simple average of the two forecasts performs better than either one of them overall, even though no average can outperform both component forecasts for any given state-date pair. This effect arises because the model and the market make different kinds of errors in different states: the model was confidently wrong in some cases, while the market was excessively uncertain in others. We propose a method for generating hybrid forecasts based on a trading bot endowed with a budget, tunable risk preferences, and beliefs inherited from the model. We also propose and conduct a profitability test that can be used as a novel criterion for the evaluation of model forecasting performance.
\end{abstract}

\thispagestyle{empty}




\newpage

\section{Introduction}

The forecasting of elections is of broad and significant interest. Predictions influence contributions to campaigns, voter turnout, candidate strategies, investment decisions by firms, and a broad range of other activities. But accurate prediction is notoriously challenging, and each approach to forecasting has significant limitations.


In this paper we compare two very different approaches to the forecasting of elections, demonstrate the value of integrating them systematically, and propose a method for doing so. One is a model-based approach that uses a clearly identified set of inputs including polls and fundamentals to generate a probability distribution over outcomes \citep{heidemanns2020}. Daily state-level forecasts based on such a model were published by \textit{The Economist} for several months leading up to the 2020 presidential election in the United States. Over the same period, a market-based approach to prediction was implemented via decentralized trading on a peer-to-peer exchange, \textit{PredictIt}, which operates legally in the United States under a no action letter from the Commodity Futures Trading Commission. Daily closing prices on this exchange, suitably adjusted, can be interpreted as probabilistic forecasts. 

Models and markets both respond to emerging information, but do so in different ways. Most models are backward-looking by construction; they are calibrated and back-tested based on earlier election cycles, and built using a set of variables that have been selected prior to the prediction exercise. New information is absorbed only when it starts to affect the input variables, for instance by making its presence felt in polling data or economic indicators. By contrast, markets are fundamentally forward-looking, and can rapidly incorporate information from novel and essentially arbitrary sources, as long as any trader considers this to be relevant. Information that affects the prices at which traders are willing to buy and sell contracts changes the market equilibrium, and hence the implied probabilistic forecasts derived from prices.

Each approach has strengths and weaknesses. New information that knowledgeable observers know to be relevant---such as debate performances, military strikes, or sharp changes in unemployment claims---can be reflected in market prices almost instantaneously, while it leads to much more sluggish adjustments in model forecasts. However, markets can be prone to herding, cognitive biases, excess volatility, and even active manipulation. Models are generally not vulnerable to such social and psychological factors, but to avoid overfitting, they need to identify a relatively small set of variables in advance (possibly extracted from a much larger set of potentially relevant variables), and this can lead to the omission of information that turns out to be highly relevant in particular instances. Market forecasts face no such constraint, and traders are free to use any inputs, including model predictions, in forming their assessments. 

We show that data from the \textit{Economist} model and the \textit{PredictIt} market for the seven months leading up to the 2020 election exhibits the following patterns. Using mean Brier scores as a measure of accuracy, the market performs significantly better than the model during the early part of the period, especially during the month of April, and somewhat worse from September onward. Averaging over the entire period and across all battleground states, the two mechanisms exhibit comparable performance. However, they generate quite different forecasts for individual states, and make errors of different kinds. For example, the model was confident but wrong in Florida and North Carolina, while the market was excessively uncertain in Minnesota and New Hampshire. This raises the possibility that a simple average of the two predictions could outperform both component forecasts in the aggregate, by avoiding the most egregious errors. We verify that this is indeed the case. Even though no average can ever outperform both components for any given state-date pair, the simple average does so in the aggregate, across all states and periods. 

A simple average of model and market forecasts is a crude approach to generating a synthetic prediction. The fact that it nevertheless exhibits superior performance to either component suggests that more sophisticated approaches that combine models and markets could be very  promising. Such hybrid forecasting can be approached from two different directions. Models could directly incorporate prediction market prices as just another input variable, in combination with polls and fundamentals, or markets could be extended to allow for automated traders endowed with budgets and risk preferences that act as if they believe the predictions of the model. Following the latter approach, we show below how a hybrid prediction market can be designed and explored experimentally. 

We also conduct a performance test, by examining the profitability of a bot that held and updated beliefs based on the model forecasts throughout the period. We show that a bot of this kind would have made different portfolio choices in different states, building a large long position in the Democratic candidate in some states, while switching frequently between long and short positions and trading much more cautiously in others. It would have made significant profits in some states and lost large sums in others, but would have made money on the whole, for a double digit return. As a check for robustness, we consider how performance would have been affected had one or more closely decided states had a different outcome. This profitability test can be used as a novel and dynamic criterion for a comparative evaluation of model performance more generally. 

\section{Related Literature}

Forecasting elections based on fundamental factors such as economic conditions, presidential approval, and incumbency status has a long history \citep{fair1978, lewis1984forecasting, abramowitz1988improved, campbell1990trial, norpoth1996time, lewis1996future}. In 2020 such models faced the difficulty of an unprecedented collapse in economic activity in response to a global pandemic, and were forced to make a number of \textit{ad hoc} adjustments \citep{abramowitz2020s, jerome2020state, desart2020long, enns2020forecasting, lewis2020political, murr2020citizen,norpoth2020primary}. While they varied widely in their predictions for the election, the average across a range of models was reasonably accurate \citep{dassonneville2020}.

Traditional models such as these typically generate forecasts with low frequency, often just once in an election cycle. More recently, with advances in computational capacity, data processing, and machine learning, models have been developed that rapidly incorporate information from trial heat polls and other sources to generate forecasts at daily frequencies. In the 2020 cycle there were two that received significant media attention, published online at \textit{FiveThirtyEight} and \textit{The Economist}; see \citet{silver2020fivethirtyeight} and  \citet{heidemanns2020} for the respective methodologies. We focus on the latter because it has greater clarity and transparency, though our analysis could easily be replicated for any model that generates forecasts for the same set of events.

Prediction markets have been a fixture in the forecasting ecosystem for decades, dating back to the launch of the pioneering \textit{Iowa Electronic Markets} in 1988 \citep{forsythe1992anatomy}. This market, like others that followed in its wake, is a peer-to-peer exchange for trading contracts with state-contingent payoffs. Buyers pay a contract price and receive a fixed payment if the referenced event occurs by an expiration date, and get no payment otherwise. The contract price is commonly interpreted as the probability of event occurrence. The forecasting accuracy of such markets has been shown to be competitive with those of opinion polls and structural models \citep{leigh2006competing, berg2008results, berg2008, rothschild2009}. 

A prediction market is a  type of crowdsourced forecasting pool that leverages a market structure, similar to a stock market, to efficiently combine judgments, provide timely responses, and incentivize truthful reporting of beliefs \citep{wolfers2004prediction}.  Forecasting crowds are not constrained by a fixed methodology or a preselected set of variables, and if suitably diverse, can include a range of ideas and approaches. Crowdsourcing forecasts have been demonstrated to be more accurate than experts' predictions \citep{satopaa2014combining}, which can be attributed to pooling and balancing diverse opinions~\citep{budescu2007aggregation}. A market can predict well even if individual participants predict poorly, through endogenous changes in the distribution of portfolios over time \citep{kets2014betting}.

The value of combining or aggregating forecasts derived from different sources has been discussed by many authors; see \citet{clemen1989combining} for an early review and \citet{rothschild2015combining} and \citet{graefe2015limitations} for more recent discussions. We propose a novel method for aggregating predictions through the actions of a trading bot that internalizes the model and is endowed with beliefs, a budget, and risk preferences. In some respects the activities of this trader resemble those of a traditional specialist or market maker, as modeled by \citet{glosten1985bid} and \citet{kyle1985continuous}, as well as algorithmic traders of more recent vintage, as in \citet{budish2015high} and \citet{baron2019risk}. But there are important differences, as we discuss in some detail below.

\section{Data}

\citet{heidemanns2020} constructed a dynamic Bayesian forecasting model that integrates information from two sources---trial heat polls at the state and national level, and fundamental factors such as economic conditions, presidential approval, and the advantage of incumbency---while making allowances for differential non-response by party identification and the degree of political polarization. The model builds on \citet{linzer2013}, and uses a variant of the \citet{abramowitz2008} fundamentals model as the basis for a prior belief. State-level forecasts based on this model were updated and published daily, and we use data for the period April 1 to November 2, 2020, stopping the day before the election.   

For the same forecasting window, we obtained daily closing price data for state markets from the \textit{PredictIt} exchange. These prices are for contracts that reference events, and each contract pays a dollar if the event occurs and nothing otherwise. For example, a market based on the event "Which party will win Wisconsin in the 2020 presidential election?" contained separate contracts for the ``Democratic'' and ``Republican'' nominees. At close of trading on November 2, the eve of the election, the prices of these contracts were 0.70 and 0.33 respectively. These cannot immediately be interpreted as probabilities since they sum to something other one.\footnote{This is a consequence of the fee structure on \textit{Predictit}, which took ten percent of all trader profits. In the absence of fees, a trader could sell both contracts and effectively purchase \$1.03 for \$1.00, making the prices unsustainable. The \textit{Iowa Electronics Market} does not charge fees and the prices of complementary contracts on that exchange preclude such arbitrage opportunities.} To obtain a probabilistic market forecast, we simply scale the prices uniformly by dividing by their sum. In this instance the market forecast for the likelihood of a Democratic victory is taken to be about 0.68. 

Traders on \textit{PredictIt} were restricted to a maximum position size of \$850 per contract. The effects on predictive accuracy of such restrictions is open to debate, but the limits do preclude the possibility that a single large trader could dominate or manipulate the market and distort prices, as was done on the \textit{Intrade} exchange during the 2012 cycle \citep{rothschild2016}. Furthermore, despite the restrictions on any individual trader's position sizes, total trading volume on this exchange was substantial. Over the seven-month period under consideration, for markets referencing outcomes in thirteen battleground states, more than 58 million contracts were traded. 

Our focus is on 13 swing states that were all considered competitive to some degree at various points in the election cycle: Arizona, Florida, Georgia, Iowa, Michigan, Minnesota, Nevada, New Hampshire, North Carolina, Ohio, Pennsylvania, Texas, and Wisconsin. We exclude states that were considered non-competitive both because there is little disagreement between the model and the market, and also because much of the disagreement that exists arises because the incentive structure created by markets distorts prices as they approach extreme levels.\footnote{Specifically, as the price of a contract approaches 1 for events that are considered very likely, contracts will have high market prices and negligible potential gain, while betting against the event will have negligible cost and high gain with low probability. This results in a demand-supply imbalance that distort prices away from underlying beliefs.}

\begin{center}
\begin{figure}[htbp]
    \includegraphics[width=\textwidth]{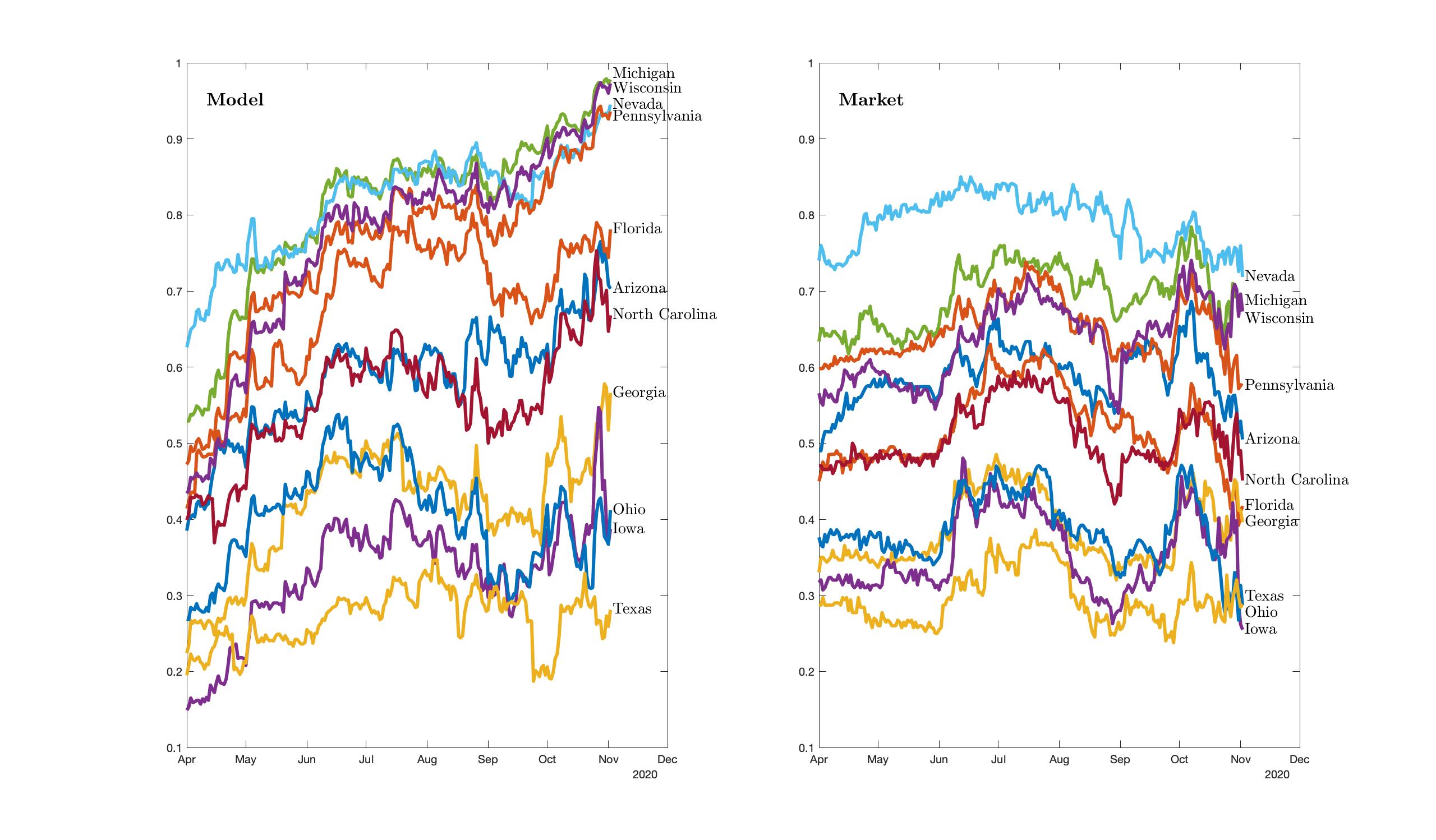}
    \caption{Probabilities of a Democratic Victory in Selected States, based on Model and Market \label{forecasts}}
  \end{figure}
\end{center}

Figure \ref{forecasts} shows the probability of a Democratic victory in eleven of the thirteen states based on the model and the market.\footnote{Two of the states---New Hampshire and Minnesota---have been dropped from the figure for visual clarity; these exhibit roughly similar patterns to the state of Nevada.} As can be seen, both the levels and trends diverge quite substantially. The model began to place extremely high likelihood (close to 100 percent) on a Democratic victory in Michigan, Wisconsin, Nevada, and Pennsylvania as the election approached, while the market forecast was much less certain, especially for Pennsylvania. 
 
\begin{center}
\begin{figure}[htbp]
    \includegraphics[width=\textwidth]{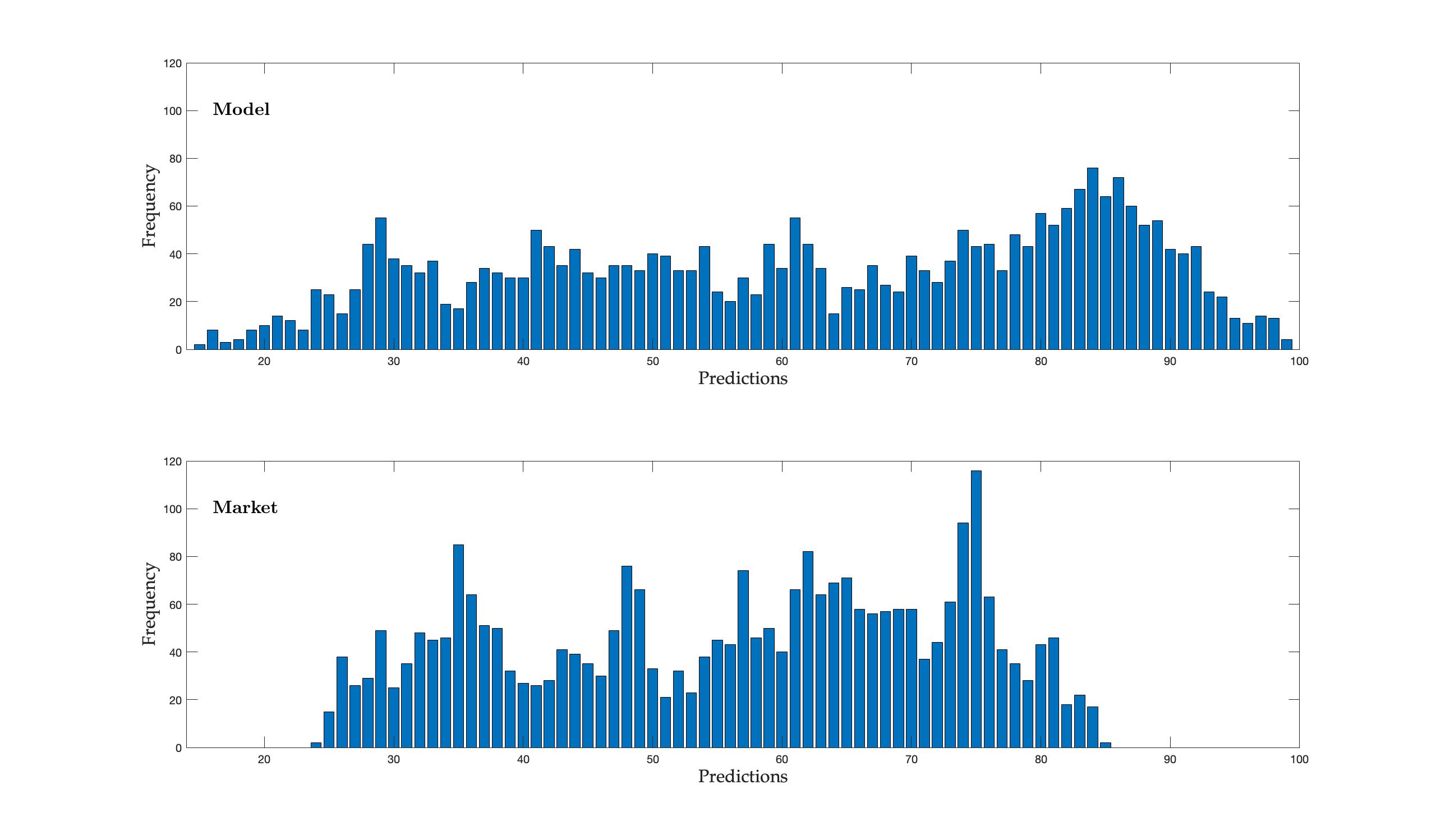}
    \caption{Frequency Distributions for Model and Market Predictions \label{frequencies}}
  \end{figure}
  \end{center}

As Figure \ref{forecasts} reveals, the model predictions span a larger subset of the probability space than the market predictions do, especially over the latter part of the period. This can be seen more clearly clearly in Figure \ref{frequencies}, which shows the frequencies with which specific probabilities appear in the data, aggregating across all thirteen states. The range of the market data is considerably more compressed.

 Given this, one might expect that at any given probability level, more states would appear at least once over the period of observation. However, this is not the case. At most seven states appear at any given prediction point in the market, and this happens just once at probability point 0.63. The model has more than 25 prediction points at which seven or more states appear, and the peak is 10 states at probability point 0.54. That is, not only does the model span a larger range of probabilities, it also has greater movement across this range for individual states. These discrepancies indicate differences in the ways in which models and markets aggregate information. The market forces updates to be frequent (because the number of traders is large) and to be small (because no individual can move prices much). Small, frequent updates have been asociated with better accuracy among forecasters \cite{atanasov2020small}.
 
\section{Performance}

A standard tool for assessing the accuracy of probabilistic forecasts is the Brier score \citep{brier1950}. In the case of binary outcomes this is just the mean squared error. Letting $p_{it}$ denote the probability assigned in period $t$ to a Democratic victory in state $i$, the score is given by 
$$ s_{it} = (p_{it} - r_i)^2, $$
where $r_i = 1$ if state $i$ is resolved in favor of the Democratic nominee and $r_i = 0$ otherwise.
  
\begin{center}
\begin{figure}[htbp]
    \includegraphics[width=\textwidth]{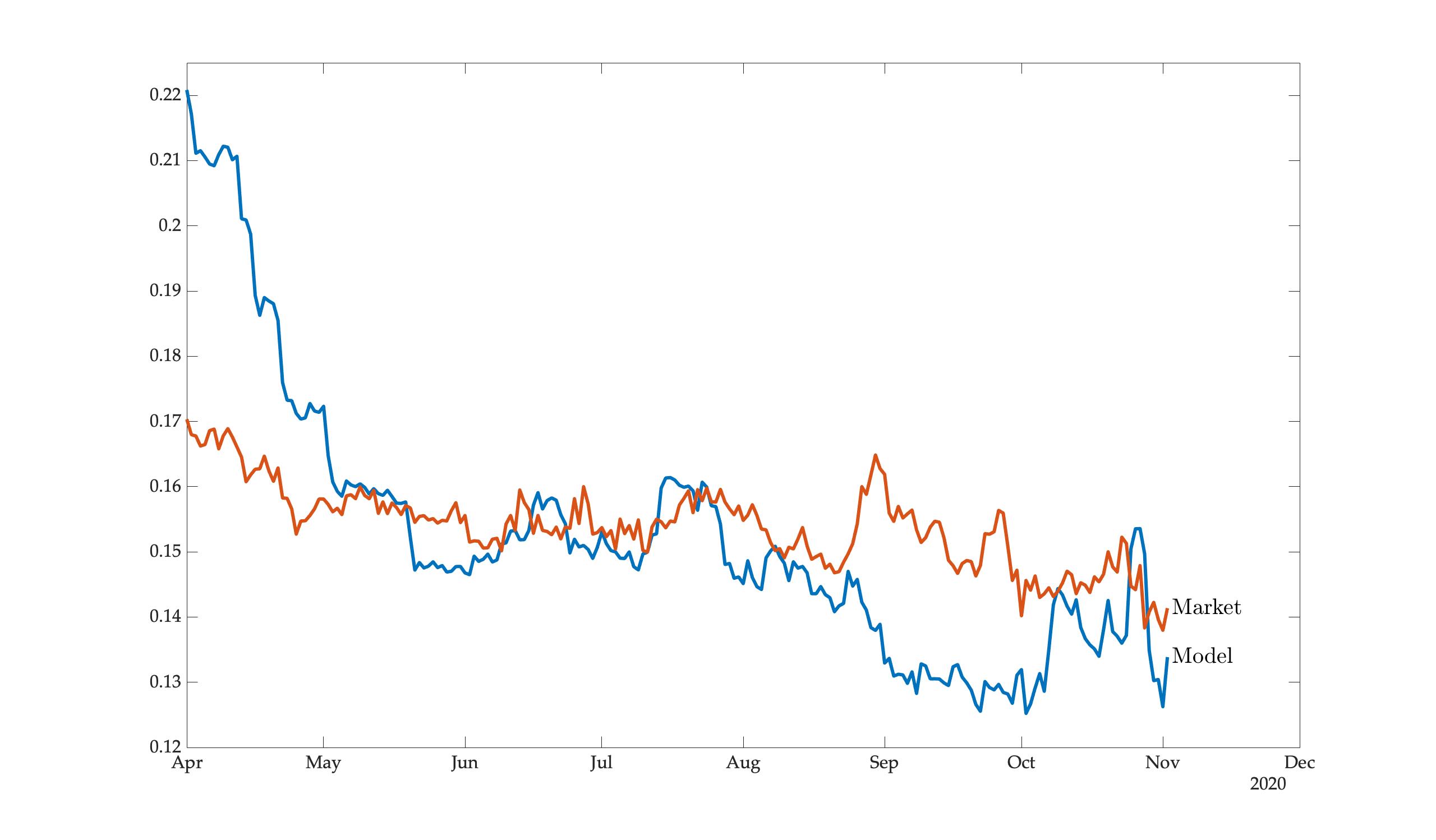}
    \caption{Mean Brier Scores for the Model and Market over Time\label{briers}}
  \end{figure}
  \end{center}

We have $n = 13$ events (corresponding to outcomes in the battleground states), for each of which there are 216 forecasts on consecutive days. Aggregating across all events we obtain a time series of the average Brier score: 
$$ \bar{s}_{t} = \frac{1}{n} \sum_{i=1}^{n} (p_{it} - r_i)^2. $$
This can be computed for models and markets separately, resulting in the  plots in Figure \ref{briers}. As we have already seen, the market forecasts were much less variable over time, both for individual states and in the aggregate. Looking at relative performance, we see that markets were significantly more accurate on average during the early part of the forecasting window. This was followed by a period of roughly comparable performance, with the model pulling ahead as the election approached.  
  
Averaging across time as well as states, we can obtain a scalar measure of overall performance for each method as follows:
$$ \bar{s} = \frac{1}{nT} \sum_{t=1}^T \sum_{i=1}^{n} (p_{it} - r_i)^2, $$
where $T = 216$ is the number of periods in the forecasting window. On this measure we find virtually identical average forecasting performance across methods:
\begin{eqnarray}
\bar{s}^{model} & = & 0.1523 \label{s1} \\
\bar{s}^{market} & = & 0.1539 \label{s2}
\end{eqnarray}
However, this conceals significant differences in performance over time as are evident in Figure \ref{briers}.

\begin{center}
\begin{figure}[h!]
    \includegraphics[width=\textwidth]{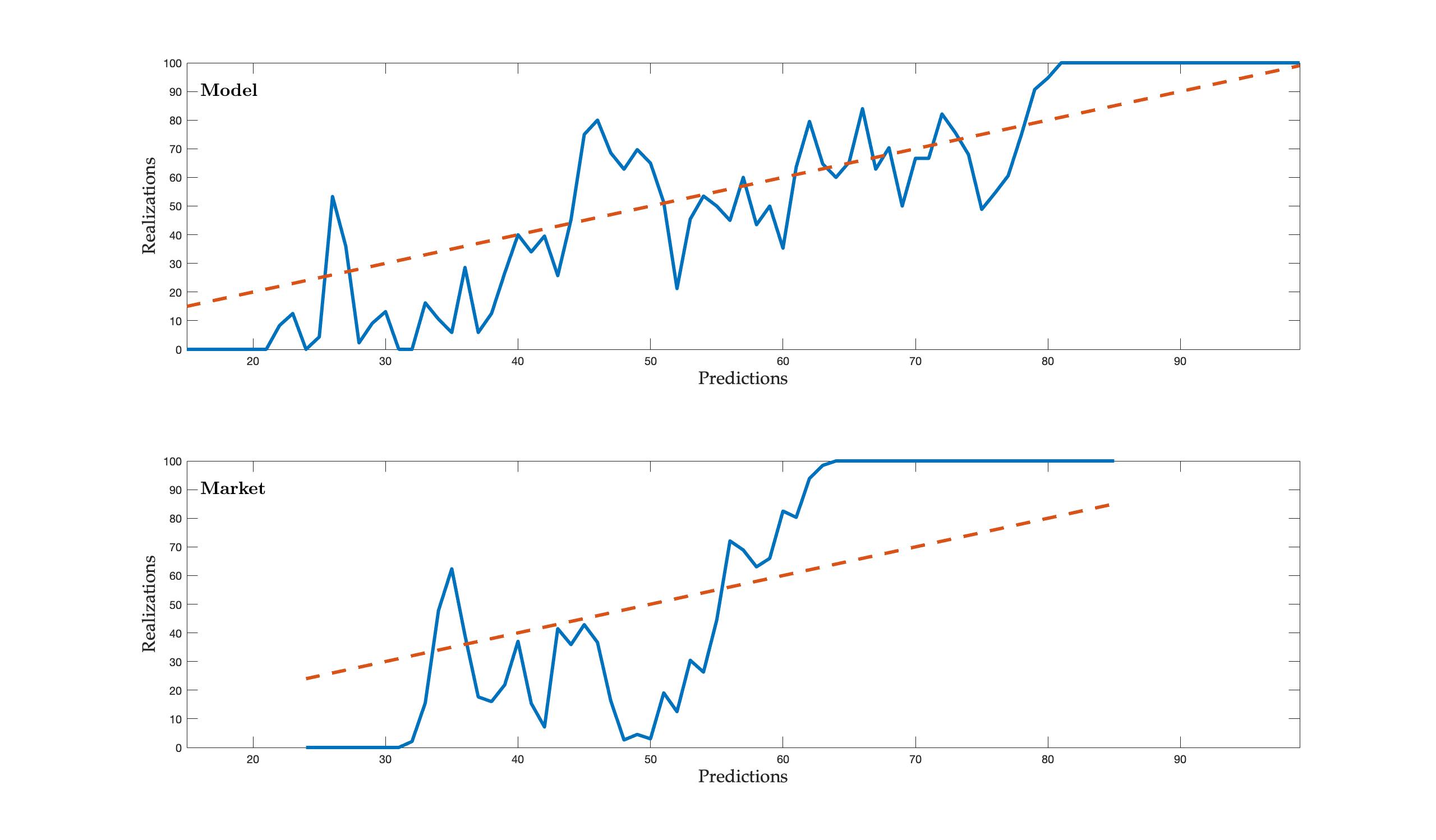}
    \caption{Calibration Curves for the Model and Market\label{calibration}}
  \end{figure}
  \end{center}

As an alternative performance measure, consider calibration curves for the model and market forecasts, shown in Figure \ref{calibration}. Again, the small number of total events being forecast leads to a lot of volatility, since the inclusion or exclusion of a single state at a prediction point can shift the realized outcome sharply. At the extreme points of the range, where only one state appears, the realizations for all forecasts have to be either zero or one. As is apparent from the figure, the model is quite well calibrated across the range of forecasts. The market forecasts span a more compressed range, as we have already seen, and the calibration appears less impressive.

\begin{center}
\begin{figure}[htbp]
    \includegraphics[width=\textwidth]{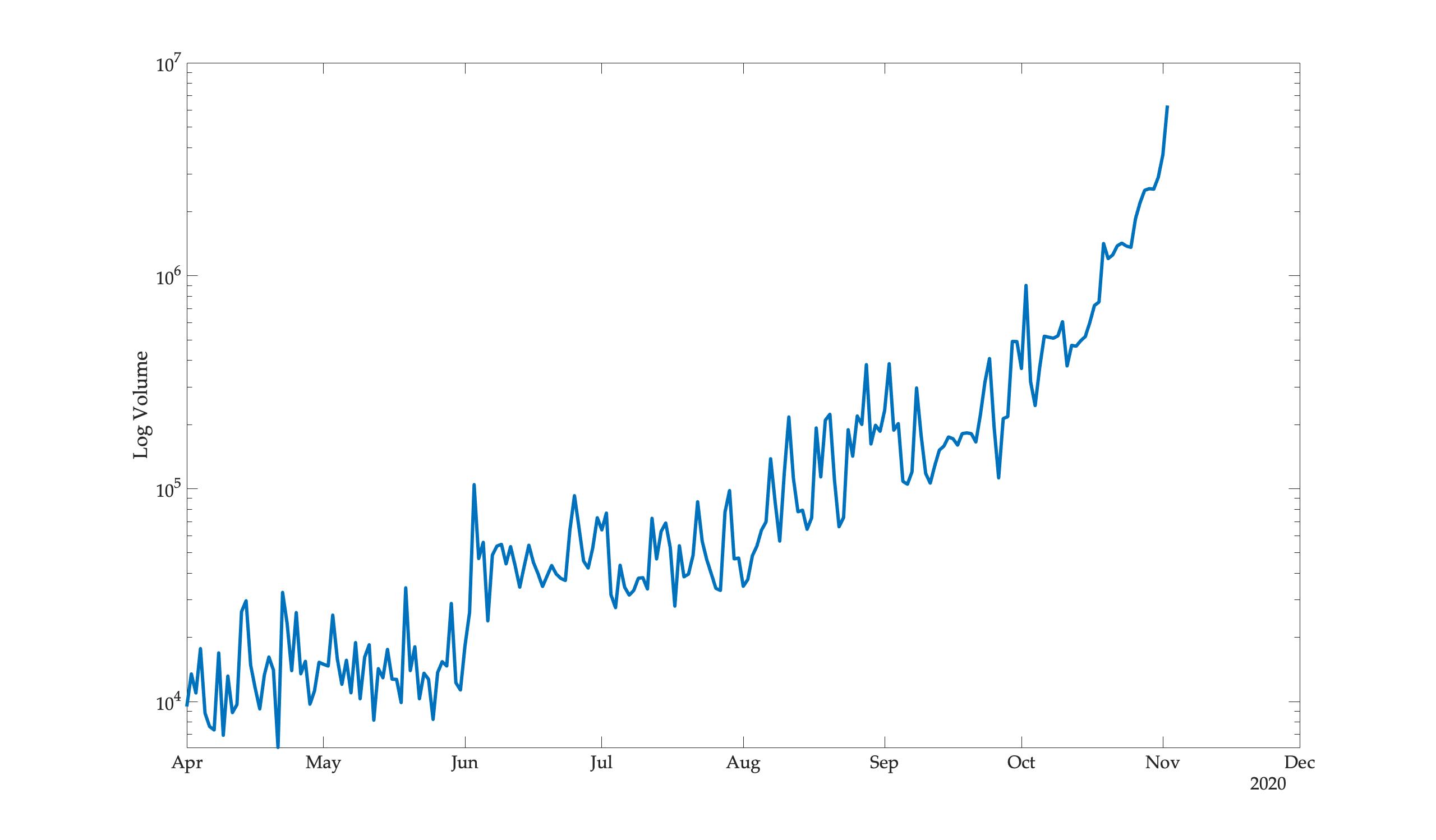}
    \caption{Trading Volume for Major Party Contracts in Thirteen Battleground States\label{volume}}
  \end{figure}
  \end{center}

One factor that may have distorted prices as the election approached was a staggering increase in volume in the wake of a sustained inflow of funds.  This can be seen in Figure \ref{volume}, which shows the total number of contracts traded daily in the thirteen battleground states over the seven month period, using a log scale. Daily volume by the end of the period was a thousand times as great as at the start. In fact, volume continued to remain extremely high even after the election was called on November 7, driven in part by traders who were convinced that the results would somehow be overturned. On November 8, for instance, there were more than 2.7 million contracts traded for six key contested states (Arizona, Georgia, Michigan, Nevada, Pennsylvania, and Wisconsin). All of these had been called for Biden but the implied market forecast for the Democrat ranged from 0.86 in Arizona to 0.93 in Nevada. This suggests that even leading up to the election, a group of traders convinced that a Trump victory was inevitable and willing to bet large sums on it were having significant price effects \citep{strauss_2021}.

Even setting aside this consideration, it is important to note that our comparison between  model and market performance ought not to be seen as definitive. Although we examine more than 2,800 predictions for each mechanism over the seven month period, they reference just thirteen events, and the scores are accordingly very sensitive to changes in outcome. For example, the model assigned a 97 percent likelihood of a Biden victory in Wisconsin on the eve of the election, and this state was decided by less than 21,000 votes, about 0.6 percent. Had there been a different outcome in Wisconsin (or indeed in Michigan, Pennsylvania, or Nevada, all of which were close, and all of which the model predicted Biden would win with probability 93 percent or higher) the model's measured performance would have been much worse. The value of examining relative performance lies in identifying systematic variation across time, the nature of errors, and the value of hybridization. We consider this next.

\section{Synthetic Forecasts}

\begin{center}
\begin{figure}[htbp]
    \includegraphics[width=\textwidth]{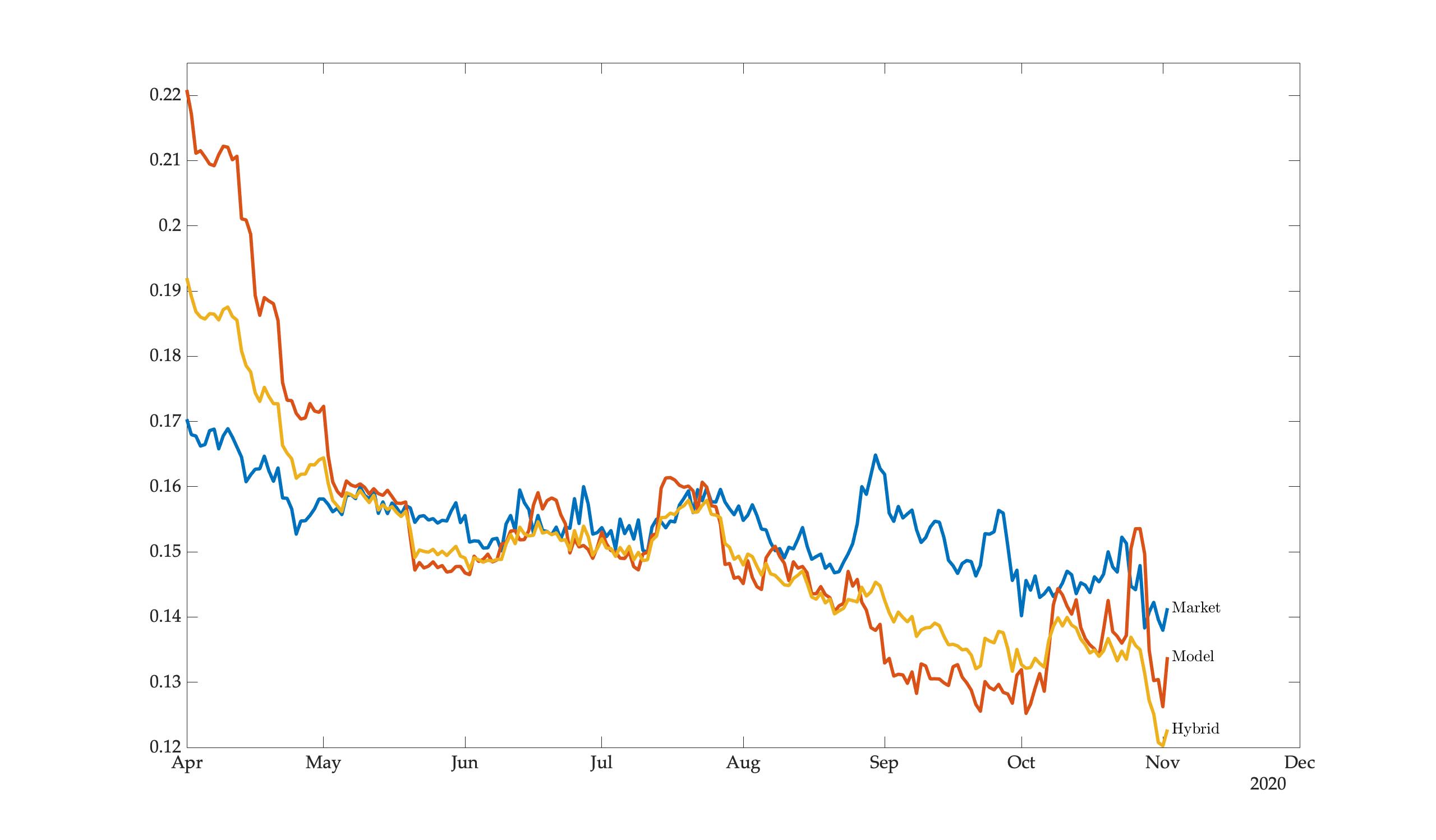}
        \caption{Mean Brier Scores for the Model, Market, and a Simple Average over Time\label{synthetic}}
  \end{figure}
  \end{center}

As a first step to demonstrating the value of combining model and market forecasts, we compute for each state-date pair the simple average estimate for the probability of a Democratic victory. Brier scores for this synthetic forecast are shown in Figure \ref{synthetic}, along with the separate model and market scores shown earlier in Figure \ref{briers}.

By construction, for any given state-date pair, the score for the synthetic forecast must lie between model and market scores; it cannot be lower than both. But this is no longer true when we average across states at any given date. In fact, as Figure \ref{synthetic} clearly shows, there are many dates on which the synthetic forecast performs better than \textit{both} model and market. Of the 216 days in the series, there are 87 on which the synthetic forecast beats both model and market, including each of the 26 days leading up to the election. Furthermore, across the entire time period and all states, the hybrid forecast received a Brier score of 
  \begin{equation}
        \bar{s}^{hybrid} = 0.1499. \label{s3}
  \end{equation}
Comparing (\ref{s3}) to (\ref{s1})--(\ref{s2}) we see that the simple average of the two component forecasts outperformed both market and model.
  
\begin{center}
\begin{figure}[htbp]
    \includegraphics[width=\textwidth]{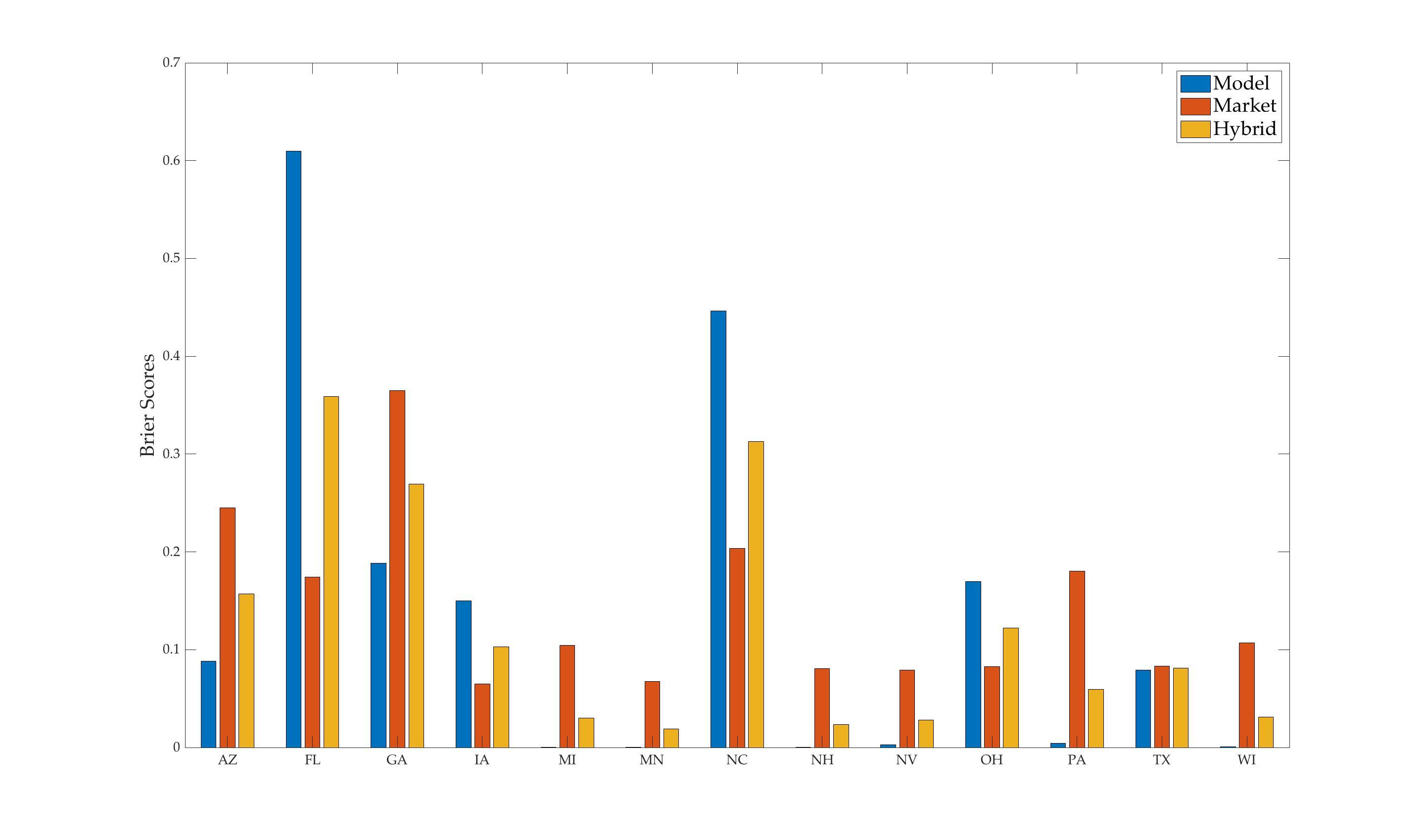}
        \caption{Brier Scores for Market, Model, and Hybrid Forecasts on November 2\label{bars}}
  \end{figure}
  \end{center}

How can a simple average of two forecasts be superior to each of the component forecasts? The reason is that the Brier score penalizes large errors, and the model and market make such errors in different ways and for different states. Consider, for example, the scores for just the November 2\textsuperscript{nd} forecasts, which are shown in Figure \ref{bars} for each of the states. For any given state the hybrid forecast Brier score necessarily lies between the two other scores. But averaging across states for this date, we get scores of 0.1414 for the market, 0.1339 for the model, and 0.1228 for the hybrid forecast. The market fared poorly in states such as New Hampshire and Minnesota, for which it did not make confident forecasts even though they ended up not being especially close. The model fared poorly in Florida and North Carolina, where it predicted Democratic wins with probabilities 0.78 and 0.67 respectively, neither of which materialized. 

To summarize, the model fared poorly in states that it confidently expected to go to Democratic nominee, and which he ended up losing in the end. The market fared poorly in a number of states where it correctly predicted that the Democrat would win, but with much lower confidence than the model. The hybrid forecast could not beat both model and market for any given state, but was able to beat both when averaging across states, by avoiding the most egregious errors.

The simple unweighted average, taken uniformly across time, is the crudest possible way of generating a hybrid forecast.\footnote{See, however, \citet{dawes_robust_1979} on the "robust beauty" of linear models and how even unsophisticated models succeed at rule-based prediction and minimizing error.} One alternative is to use prediction market prices as direct inputs in the model, along with information from trial-heat polls and fundamentals. A different approach, which we describe next, consists of constructing markets in which a model is represented by a virtual trader. 

\section{Hybrid Prediction Markets}


Algorithmic trading is a common feature of modern financial markets, where low latency can be highly rewarding \citep{budish2015high, baron2019risk}.\footnote{Algorithmic trading was also extremely common on the \textit{Intrade} prediction market \citep{rothschild2016}, even though it is disallowed on \textit{PredictIt}. } In this sense markets for stocks and derivatives are already hybrid  markets. But  algorithms in these settings operate as part of an organic trading ecosystem, competing with each other and more traditional strategies implemented by fund managers and retail investors. In order to construct a hybrid prediction market that can be tuned and tested experimentally, we propose a somewhat different approach. 

Given a model that generates forecasts updated with relatively high frequency, one can insert into the market a trading bot that acts as if it believes the model forecast. In order to do so, the bot has to be endowed with a budget and preferences that exhibit some degree of risk aversion. These parameters can be tuned in experimental settings to examine their effects on forecasting accuracy. The bot posts orders to buy and sell securities based on its beliefs (derived from the model), its preferences, and its existing portfolio, which consists of cash and various asset positions. These orders either trade immediately or enter the order book, providing opportunities for human traders to bet on or against the referenced event if their own assessments of its likelihood differ. And through the process of trading, the cash and asset holdings of the bot evolve over time, affecting the prices and sizes of its orders. 

To be precise, consider a model that generates a joint distribution over the electoral outcomes in $m$ jurisdictions, where each jurisdiction has $n$ candidates on the ballot.\footnote{There is no loss of generality in assuming that all jurisdictions have the same number of candidates, since any with fewer than $n$ candidates can be assigned notional candidates who win  with zero probability.} Let $\textbf{S}$ denote an $n \times m$ matrix that represents an outcome realization, with typical element $s_{ij} \in \{0,1\}$. Columns $\textbf{s}_j$ of $\textbf{S}$ correspond to jurisdictions, and $s_{ij} = 1$ indicates that candidate $i$ is the (unique) winner in jurisdiction $j$. Let $\Omega$ denote the set of all possible electoral outcomes, and let $p: \Omega \to [0,1]$ denote the probability distribution over these outcomes generated by the model. 

Now suppose that for each jurisdiction there exists a prediction market listing $n$ contracts, one for each candidate. In practice, there will be distinct prices for the sale and purchase of a contract, the best bid and best ask prices. We return to this issue in Section \ref{secmm} below, but assume for the moment that there is no such difference, and that each contract is associated with a unique price at which it can be bought or sold.\footnote{For battleground state markets on \textit{PredictIt} the bid-ask spread was typically just a penny, which large numbers of contracts available for purchase and sale at most times.} Let $\textbf{Q}$ denote a set of prices in these markets, with typical element $q_{ij} \in [0,1]$. Here $q_{ij}$ is the price at which one can purchase or sell a contract that pays a dollar if candidate $i$ wins in jurisdiction $j$, and pays nothing otherwise. Columns $\textbf{q}_j$ of $\textbf{Q}$ contain the prices for each of the $i$ contracts in jurisdiction $j$.

Next consider a virtual trader or bot that represents the model in the market, in the sense that it inherits and updates beliefs over the set of outcomes based on the output of the model. At any given point in time, this trader in this market will be holding a portfolio $(y,\textbf{Z})$, where $y$ is cash and $\textbf{Z}$ is a $n \times m$ matrix of contract holdings. Element $z_{ij} \in \R$ is the number of contracts held by the trader that reference candidate $i$ in jurisdiction $j$. It is possible for $z_{ij}$ to be negative, which corresponds to a short position: if candidate $i$ wins in jurisdiction $j$ then a trader with $z_{ij} < 0$ will pay (rather than receive) a dollar per contract. Buying a contract referencing candidate $i$ in jurisdiction $j$ lowers cash by $q_{ij}$ and raises $z_{ij}$ by one unit, while selling such a contract raises cash by $q_{ij}$ and lowers $z_{ij}$ by one unit. Let $\textbf{z}_j$ denote  column $j$ of $\textbf{Z}$, the contract holdings in the market corresponding to jurisdiction $j$.

If the outcome is $\textbf{S} \in \Omega$ when all contracts are resolved, the terminal cash value or wealth resulting from portfolio  $(y,\textbf{Z})$ is 
$$ w = y + \sum_{j \in M} \textbf{s}'_{j}\textbf{z}_{j},$$
where $M = \{1,...,m\}$ is the set of jurisdictions. A risk-neutral trader would choose a portfolio that maximizes the expected value $E(w)$ of terminal wealth, given her beliefs $p$ and the market prices $\textbf{Q}$, subject to solvency constraints that are discussed further below. A risk-averse trader, instead, will maximize expected utility, given by 
$$ E(u) = \sum_{\textbf{S} \in \Omega} p(\textbf{S}) u \left(y + \sum_{j \in M}  \textbf{s}'_{j}\textbf{z}_{j}  \right),$$
where $u: \R_+ \to \R$ is strictly increasing and concave. 

 Given a starting portfolio $(y,\textbf{Z})$, beliefs $p$, preferences $u$, and prices $\textbf{Q}$, let $\textbf{X}$ denote an $n \times m$ matrix of trades, where $x_{ij} \in \R$ is the (possibly negative) number of contracts purchased that reference candidate $i$ in jurisdiction $j$. Column $j$ of $\textbf{X}$, which registers trades involving jurisdiction $j$, is denoted $\textbf{x}_j$. The set of trades will be chosen to maximize
\begin{equation} \label{eugen}
E(u) =  \sum_{\textbf{S} \in \Omega} p(\textbf{S}) u \left( y + \sum_{j \in M} \left( \textbf{s}'_{j}(\textbf{z}_{j} + \textbf{x}_{j}) - \textbf{q}'_{j}\textbf{x}_{j} \right) \right).
\end{equation}
If $x_{ij} > 0$ then contracts are purchased, $z_{ij}$ rises by this amount, and $y$ falls by the cost of the contracts. If $x_{ij} < 0$ then contracts are sold, $z_{ij}$ falls by this amount, and $y$ rises by the cost of the contracts. 

A trading bot programmed to execute transactions in accordance with the maximization of (\ref{eugen}) will trade whenever there is a change in model output $p$ or in market prices $\textbf{Q}$. Any such trades must be consistent with solvency even in the worst case scenario.\footnote{Most prediction markets, including PredictIt, operate under a 100 percent margin requirement to ensure contract fulfillment with certainty, so that there is no counter-party risk or risk borne by the exchange itself.} That is, the trading bot chooses $\textbf{X}$ to maximize (\ref{eugen}) subject to the constraint:
$$ y + \min_{\textbf{S} \in \Omega} \sum_{j \in M} \left( \textbf{s}'_{j}(\textbf{z}_{j} + \textbf{x}_{j}) - \textbf{q}'_{j}\textbf{x}_{j} \right) \ge 0.$$
We next consider whether a bot trading daily in accordance with the model beliefs and a particular specification for preferences $u$, facing actual prices on \textit{PredictIt}, would have made money.

\section{A Profitability Test}

The analysis in the previous section can be put to use to develop an alternative measure of model performance, both relative to the market and to other models. Specifically, one can ask the hypothetical question of whether a bot endowed with model beliefs would have been profitable over the course of the forecasting period, when matched against the actual market participants. And one can use the level of profits or losses to compare different models, and obtain a performance criterion in which early and late forecasts are treated differently in a systematic and dynamic manner. Profitability of a model is also a necessary condition for survival, without which the model could not have a significant and persistent effect on market prices. If the model makes losses consistently its budget will diminish endogenously and it will become negligible relative to the rest of the market. Therefore only a profitable model can yield hybrid forecasts that differ meaningfully from market forecasts.

In order to implement this idea, we endowed a trading bot with a cash position $y = \$1000$ and a contract position $z = 0$. Each market has two contracts referencing the same event and we take the average of daily closing prices and allow the bot to meet all demand at this price. In practice, the price available to the bot could be slightly better as it could pick the better-priced contract or slightly worse from the bid-ask spread. Given the bot budget, its trades have a negligible effect on the order book. Additionally, we need to specify preferences for the bot. A useful class of functions for which the degree of risk aversion can be tuned for experimental purposes is that exhibiting constant relative risk aversion (CRRA):

\begin{equation*}
u(w) = \begin{cases} \frac{1}{1-\rho} w^{1-\rho} , & \mbox{if } \rho \ge 0, \; \rho \neq 1 \\ \log (w), & \mbox{if } \rho = 1 \end{cases}
\end{equation*}
where $\rho$ is the degree of relative risk-aversion. We report results below for the case of log utility. 
  
One piece of information that is unavailable to us is the complete daily probability distribution $p(S)$ over the set of possible outcomes, where $|S| = 2^{13} = 8192$. What we have, instead, are the daily marginal distributions for each state. Since state outcomes cannot be treated as independent, this is a significant loss of information. Nevertheless, we can show that a bot operating with this  limited information would have been  profitable overall, despite making losses in several states.

\begin{center}
\begin{figure}[h!]
    \includegraphics[width=\textwidth]{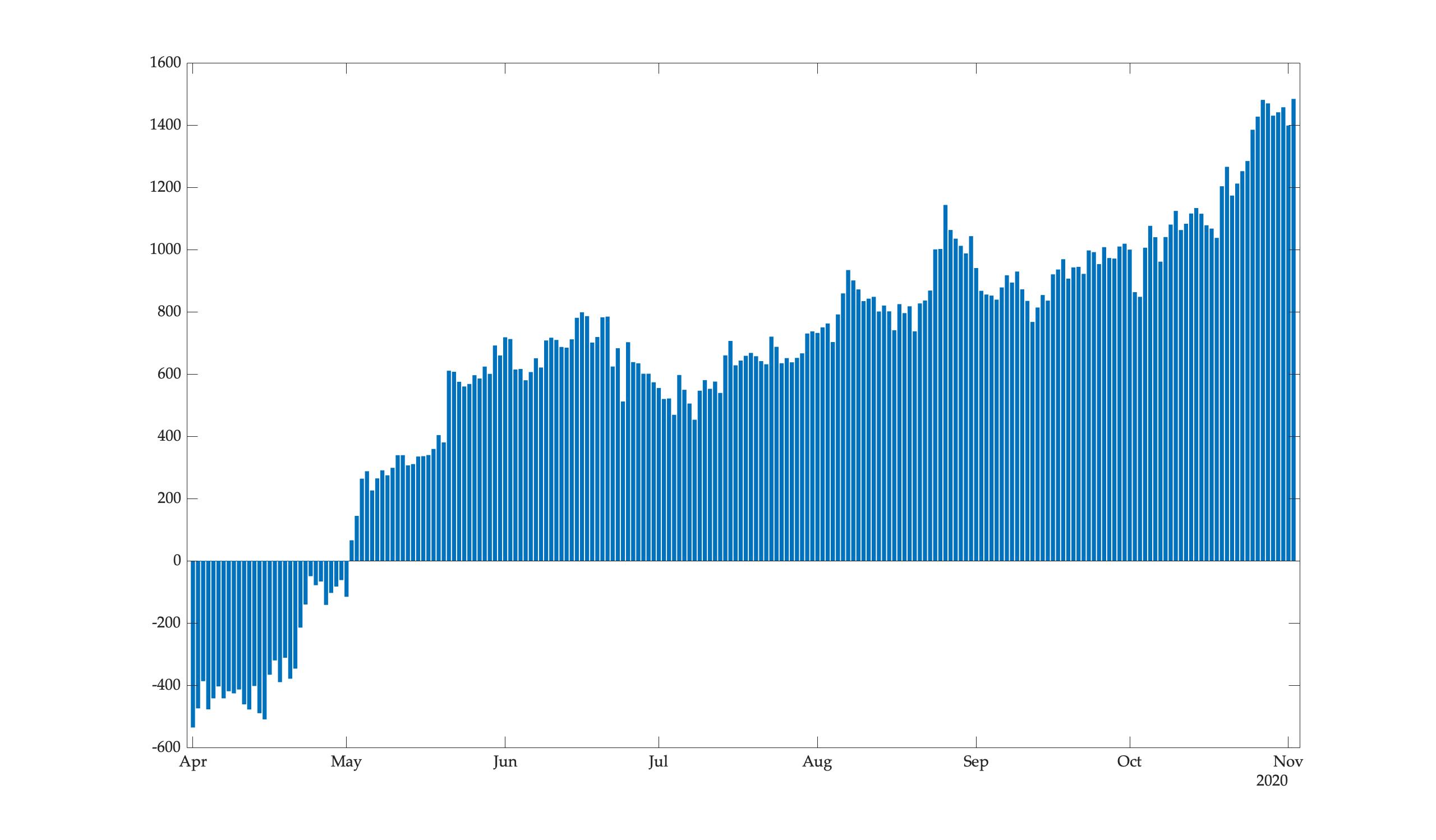}
        \caption{Evolution of Contract Holdings for Wisconsin\label{wisconsin}}
  \end{figure}
  \end{center}

We first illustrate this for the state of Wisconsin. Negative values of $z$ correspond to bets against the Democratic nominee. The evolution of $z$ over time for Wisconsin is shown in Figure \ref{wisconsin}. Initially the market considered a Biden victory to be more likely than the model did, so the bot entered a short position (negative $z$) in this contract. This started to change towards the end of April, and the bot started to cover this short position by buying contracts, eventually ending up with a long position (positive $z$) that grew unevenly but substantially over time until the election.

There are several fluctuations and interruptions in this growth, reflecting changes in beliefs and prices that warranted a less aggressive posture. But by the eve of the election the bot held less than \$90 in cash, along with almost 1,500 contracts. This portfolio had a market value of almost 1,100 on November 2. That is, the bot could have sold all contracts at the then-prevailing market price and ended up with a ten percent return without taking any risk into election day, insulated completely from the eventual outcome. These profits were generated by buying and holding contracts that had subsequently risen in price, or by buying and selling along the way. But the bot did not liquidate its position, and once the contracts were resolved, ended up with a cash position of \$1,574 for a 57\% return. Had Biden lost Wisconsin this would have resulted in a substantial loss, exceeding 90\%. 

\begin{center}
\begin{figure}[htbp]
    \includegraphics[width=\textwidth]{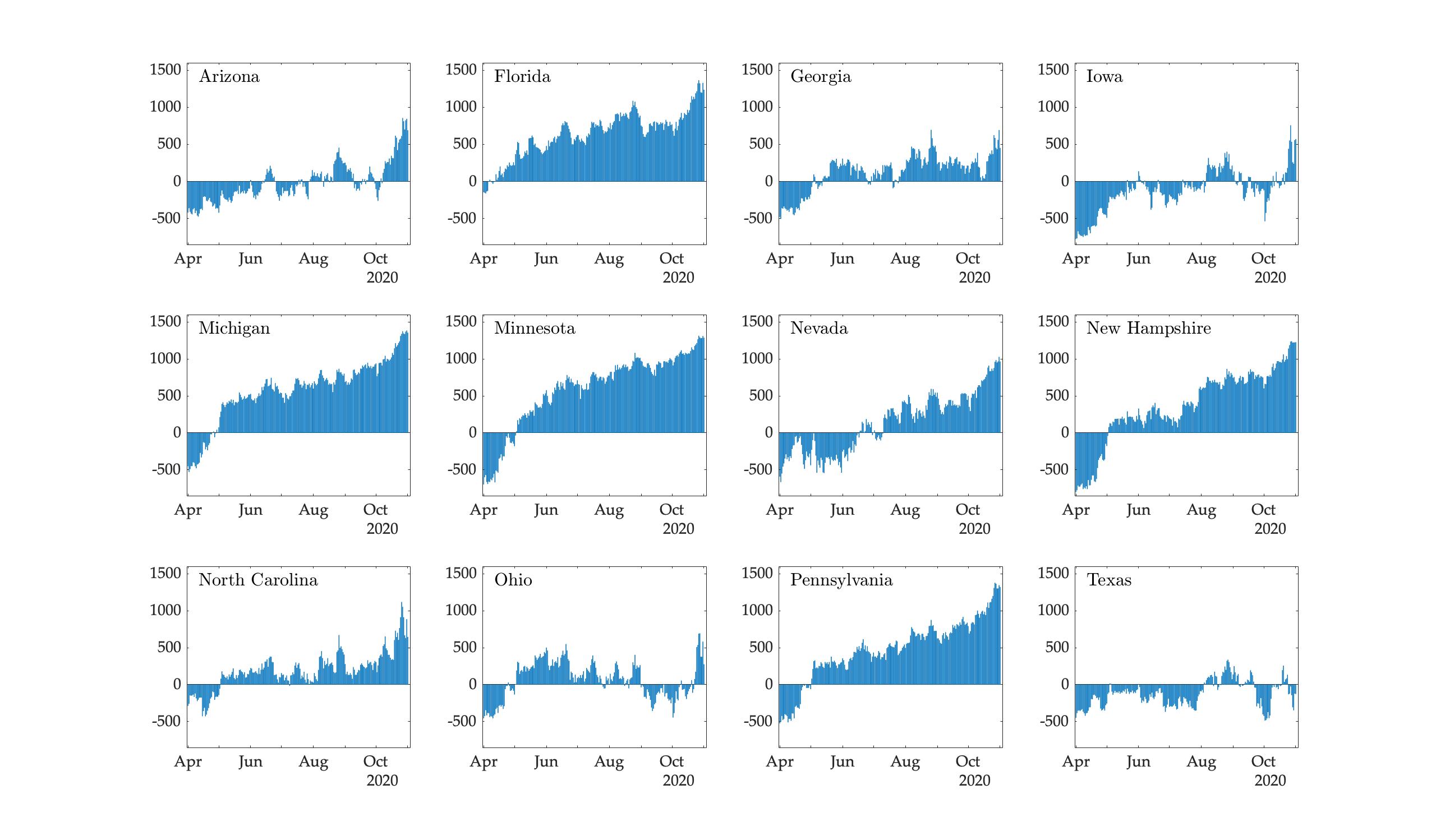}
        \caption{Bot Contract Holdings for 12 battleground States.\label{portfolios}}
  \end{figure}
  \end{center}

Repeating this analysis for the other twelve states, we get the paths shown in Figure \ref{portfolios}. In all cases the bot began by betting against a Democratic victory but quickly changed course in many states. Several states follow the same pattern as Wisconsin, in that a large positive position was built up over time. These states resulted in significant profits (Michigan, Minnesota, New Hampshire, Pennsylvania) or significant losses (Florida). There are also some states in which trading was much more cautious, because model forecasts and market prices were not far apart. And in many states contract holdings switched often from short to long and back again.

  Additional details for all states, including final portfolios, terminal payoffs, profits, and rates of return, may be seen in Table \ref{profits}. Only in Texas did the bot hold a short position in a Democratic victory on the even of the election; elsewhere the model favored the Democrat to a greater degree than the market. The market value of all portfolios on the eve of the election was just slightly greater than the starting value of \$13,000. That is, if the bot had liquidated all positions immediately prior to resolution it would have basically broken even. But the eventual payoff was much larger, with a profit exceeding \$2,000 and a 16\% return, as profits in some states outweighed losses in others. In general the bot made money where the Democrat won and lost where the Republican prevailed, with the single exception being Texas.\footnote{In Texas the cash position exceeded the value of the portfolio on election day, because the contract position was negative; had Biden won Texas the bot would have had to pay out about \$37 and would have ended up with a loss in this state.}

\begin{table}[!h]
 \begin{center}
\small
\begin{tabular}{lrrrrrr}
State          & \multicolumn{1}{c}{Cash} & \multicolumn{1}{c}{Contracts} & \multicolumn{1}{c}{Value} & \multicolumn{1}{c}{Payoff} & \multicolumn{1}{c}{Profit} & \multicolumn{1}{c}{Return} \\ \hline
Arizona        & \$566.45   & 748.55    & \$944.35     & \$1,314.99     & \$314.99   & 31\%  \\
Florida        & \$339.96   & 1351.73   & \$904.28     & \$339.96       & --\$660.04  & --66\% \\
Georgia        & \$743.54   & 735.23    & \$1,034.72   & \$1,478.76     & \$478.76   & 48\%  \\
Iowa           & \$855.20   & 722.99    & \$1,039.50   & \$855.20       & --\$144.80  & --14\% \\
Michigan       & \$68.29    & 1383.63   & \$1,004.27   & \$1,451.92     & \$451.92   & 45\%  \\
Minnesota      & \$54.98    & 1305.55   & \$1,021.09   & \$1,360.53     & \$360.53   & 36\%  \\
Nevada         & \$190.84   & 1094.15   & \$976.93     & \$1,284.99     & \$284.99   & 28\%  \\
New Hampshire  & \$72.74    & 1274.46   & \$984.85     & \$1,347.20     & \$347.20   & 35\%  \\
North Carolina & \$607.15   & 880.03    & \$1,004.03   & \$607.15       & --\$392.85  & --39\% \\
Ohio           & \$904.82   & 669.22    & \$1,096.97   & \$904.82       & --\$95.18   & --10\% \\
Pennsylvania   & \$138.91   & 1359.79   & \$921.43     & \$1,498.70     & \$498.70   & 50\%  \\
Texas          & \$1,027.13 & --36.95    & \$1,016.47   & \$1,027.13     & \$27.13    & 3\%   \\
Wisconsin      & \$89.92    & 1484.09   & \$1,088.83   & \$1,574.01     & \$574.01   & 57\%  \\ \hline
Total          &            &           & \$13,037.72  & \$15,045.36    & \$2,045.36 & 16\% \\ \hline
\end{tabular}
\caption{Terminal portfolios, payoffs, and profits in battleground states.\label{profits}}
  \end{center}
\end{table}
\normalsize

The payoffs and profits shown in Table \ref{profits} are based on the actual election outcomes. As a check on robustness, one might ask how the model would have performed if one or more close states had been decided differently. There were three states that were especially close, decided by less than one percentage point: Georgia (0.24 percent), Arizona (0.31 percent) and Wisconsin (0.62 percent). All three were won by the Democrat. Table \ref{robustness} shows the payoffs and profits that would have accrued to the bot had one or more of these been won by the Republican instead. In each case profits would have been lower, since the bot was betting on a Democratic victory in all three states. If any one of these states had been decided differently, profits would still have been positive. However, if the Democrat had lost Wisconsin along with one or both of Georgia and Arizona, it would have made a loss overall, largely because the model's very high confidence that the Democrat would win Wisconsin led the bot to accumulate a substantial number of contracts.
 
The hypothetical profits in Table 2 provide a measure of robustness by identifying the number of close states (based on some specified threshold) that would have to flip in order for the model to incur a loss. In this case the critical number or "robustness diameter" is two. For a more granular robustness measure, one could ask how many votes would be required to flip in order for the bot to lose money, and in this case the critical number is 31,141 votes (in Arizona and Wisconsin). In comparing the performance of two or more models against the market, one can consider both profitability (which depends in part on pure chance) and the robustness diameter in this way.\footnote{Note, however, that the risk preferences assigned to the bot will affect the aggressiveness of trading, position sizes, profitability, and the robustness diameter.}


 \begin{table}[h]
   \begin{center}
\small
\begin{tabular}{lccrr}
Flipped State(s)   & Margin & Payoff      & \multicolumn{1}{c}{Profit} & \multicolumn{1}{c}{Rate}    \\ \hline
Georgia            &11,780         &\$14,310.14 & \$1,310.14 & 10.08\% \\
Arizona            &10,458         &\$14,296.82 & \$1,296.82 & 9.98\%  \\
Wisconsin          &20,683         &\$13,561.27 & \$561.27   & 4.32\%  \\
Georgia, Arizona   &22,238          &\$13,561.60 & \$561.60   & 4.32\%  \\
Georgia, Wisconsin &32,463          &\$12,826.05 & --\$173.95  & --1.34\% \\
Arizona, Wisconsin &31,141          &\$12,812.73 & --\$187.27  & --1.44\% \\
Georgia, Arizona, Wisconsin &42,921 & \$12,077.51 & --\$922.49  & --7.10\% \\ \hline
\end{tabular}
\caption{Hypothetical payoffs and profits if the closest states had been decided differently.\label{robustness}}
\end{center}
\end{table}

The profitability test explored here may be improved by using the complete model-derived probability distribution over outcomes rather than just the marginal distributions for each state, and by allowing cash to flow freely across markets. This offers the possibility of evaluating a model based on its deep structure, including correlations across states, rather than just on the derived marginal distributions for each state. 
It therefore provides a new and dynamic evaluation criterion that can be used to compare models with each other, and to assess the value of hybrid forecasting more generally.

  
\section{The Model as Market Maker} \label{secmm}

To this point we have not distinguished between bid and ask prices, which will typically differ and cause the purchase price to be slightly higher that the sale price. We have also assumed that any demands or supplies can be met at prevailing market prices. But in the absence of any prices at which trading can occur, the bot representing the model can post prices at which it would be willing to trade, thus providing liquidity to the market. This is important for the development of experimental markets which are not already populated with a large number of active traders. 

As an example consider a model that forecasts a single event with $n \ge 2$ options or bins, each of which has a corresponding contract. That is, if the outcome is in bin $i$ then that contract pays out while the others expire worthless.  Suppose that a  model generates a probability distribution $p^* = (p^*_1,...,p^*_n)$ over the possible outcomes. At any point in time, the trading bot will have a cash position $y \ge 0$ and asset position $z = (z_1,...,z_n)$, where each $z_i$ can be positive or negative (representing a long or short position). The bot is endowed with a utility function $u(w)$ where $w$ is the cash holding after resolution. This is uncertain at the time of trading, and the bot places orders in order to maximize expected utility as before. 

Now consider any particular bin or outcome interval $i$. We need to find the highest price at which the bot will be willing to buy shares, and the lowest prices at which it will sell. We also need to compute the quantities posted at those prices. Suppose that at any hypothetical price $p_i$ in market $i$ the bot places an order to buy $x_i$ shares in this market. Then, given the beliefs $p^*$, its expected utility is 
\begin{equation} \label{eu}
E(u) =  p^*_i u(z_i + x_i + y - p_i x ) + \sum_{j \neq i} p^*_j u(z_j + y - p_i x_i).
    \end{equation}
Maximizing this yields a demand $x(p_i)$ for any market $i$ and price $p_i$. Let $p^b_i$ and $p^s_i$ denote the highest price at which $x(p_i) > 0$ and the lowest price at which $x(p_i) < 0$ respectively.  Then bot places orders to buy $x_i(p^b_i)$ shares at price $p^b_i$ orders to sell $x_i(p^s_i)$ shares at price $p^s_i$ in each market $i$. Note that all buy and sell prices are functions of $(y,z)$ and will change each time the bot is able to buy or sell any contracts. 

The bot is effectively acting as a market marker, as in \citet{glosten1985bid} and \citet{kyle1985continuous}, but instead of making inferences from the trading activity of counterparties, it derives and updates its beliefs based on the model. It acts as an informed trader rather than an uninformed market maker, and is willing to take and hold large directional positions. 

To illustrate, consider forecasting problem with three options or bins, and a  model that assigns probability $p^* = (0.3, 0.5, 0.2)$. The problem could involve discrete outcomes that can take one of three values, or continuous outcomes that can fall into three specified ranges. The model forecast is represented in the market by a bot having initial cash endowment $y = \$1000$ and preferences given by $u(w) = \log(w)$. The bot initially holds no contracts, so $z = (0,0,0)$. 

To begin with, the bot places six orders, one to buy and one to sell in each of the three bins, resulting in the following order book:

\begin{table}[h!]
\begin{center}
\begin{tabular}{c|cccc}
Bin & Bid Price & Bid Quantity & Ask Price & Ask Quantity \\ \hline
1 & 0.29  & 48  & 0.31  & 46 \\ 
2 &  0.49   & 40 &    0.51 &  40 \\
3 &  0.19 &   64   &  0.21 &   60 \\ 
\end{tabular}
\end{center}
\end{table}%

\noindent This is what human traders see when they view the market. For each bin the prices posted by the bot straddle the beliefs generated by the model, though this can change as cash and asset positions are altered through the process of trading, as we shall see momentarily. The bid-ask spreads are tight, and there is plenty of liquidity available for humans who wish to trade. 

Now suppose that the offer to buy 48 shares at 0.29 is met (a human trader chooses to bet against the event corresponding to Bin 1, paying 0.71 per contract for 48 contracts, each of which will pay out a dollar if this event does \textit{not} occur). Then the asset position of the bot goes to $z = (48,0,0)$ and its cash position goes to $1000 - 0.29(48) = 986.08$. At this point he bot cancels all remaining orders and starts afresh, placing six new orders by maximizing (\ref{eu}) with the current portfolio as starting point. As a result, prices and quantities change in \textit{all} markets, not just in the market in which the transaction occurred, resulting in the following order book:

 \begin{table}[h!]
\begin{center}
\begin{tabular}{c|cccc}
Bin & Bid Price & Bid Quantity & Ask Price & Ask Quantity \\ \hline
1 & 0.28  & 51  & 0.30  & 47 \\ 
2 &  0.50   & 28 &    0.51 &  11 \\
3 &  0.20 &   17   &  0.21 &   42 \\ 
\end{tabular}
\end{center}
\end{table}%

Note that the bot is now willing to sell shares in the first bin at price 0.30, even though this exactly matches its belief. This is because it can secure a reduction in the risk of its portfolio, without lowering expected value. In fact, if it continues to accumulate shares in the first bin, it will offer to sell at prices below 0.30, lowering expected return but also lowering risk.

A naive observer of this market will see a bot acting as a market maker, placing orders on both sides of the market, and canceling and replacing these whenever any order is met. But the strategy implemented by the bot is not that of a traditional market maker; it is maximizing expected utility subject to model-based beliefs about the various outcomes. In fact, when the model is updated the bot's imputed beliefs $p^*$ change, and it will respond by canceling and replacing all orders, possibly resulting in the immediate execution of trades against humans. And human traders can trade against each other, or post limit orders that compete with those posted by the bot.\footnote{The hybrid prediction market described here is accordingly quite different from one in which an automated market maker is a counterparty to \textit{all} trades, and sets prices based on a potential function \citep{hanson2003combinatorial,chen2007utility,sethi2016belief}.}

A market of this kind could easily be implemented experimentally, with bot preferences and budgets adjusted to test for effects on predictive performance. And unlike some existing experimental exchanges such as the \textit{Iowa Electronic Markets}, there will be substantial liquidity available to human traders at all times. No matter what the model forecast or trading history, at least one side of the order book will always be populated. As a result the spreads between bid and ask prices will be narrower, providing greater incentives for market participation. 

Social influence can be beneficial or harmful to predictive accuracy \citep{rothschild_somethings_2016,toyokawa_social_2019}. Our proposed hybrid prediction market aims to harness beneficial social influence by anchoring trader beliefs on objective, model-based predictions. Through this process the model forecasts enter market prices, integrated with the beliefs of human traders. Over time, the bot's influence on market prices will grow if it makes money, and shrink if it loses. This is as it should be---a model that predicts poorly should have diminishing influence on the hybrid forecast over time.

\section{Discussion}

The  analysis in this paper ought not to be seen as a meaningful horse race between models and markets, since all the forecasts reference just thirteen closely connected and correlated events, and a different outcome in a couple of states could have resulted in a very different evaluation of performance. Instead, our primary contribution should be seen as the development of a particular approach to hybrid forecasting that can be implemented, calibrated and tested empirically. It is important to be attentive to concerns about external validity, however, since real-world markets have recently been prone to high volatility arising from the spread of misinformation or coordinated attempts to inflate prices \citep{strauss_2021,phillips2021}.
 
 Models extrapolate from available data in a manner that is potentially objective, transparent, and consistent. They are designed to produce good average calibration across the probability space.  Markets aggregate real-time human judgement and provide agility and versatility. They balance a pool of knowledge, and are able to extrapolate from sparse data further from resolution. We have shown that even a crude simple average of these methods can outperform each one of them when examining data from the 2020 US presidential election. The lesson here is that hybrid forecasting methods are well worth developing. 


 This can be done in a number of ways. The most obvious is to simply construct a simple or weighted average, as in \citet{rothschild2015combining}, which has been found to be quite effective relative to more complex aggregation procedures \citep{clemen1989combining, graefe2015limitations}. Along similar lines, one could include prediction market prices as inputs into existing models, along with trial-heat polls and fundamental factors, calibrated based on historical data. Alternatively, one could make model outputs available to human forecasters to use as they see fit when soliciting their beliefs about events \citep{abeliuk2020}. Yet another possibility involves tapping the wisdom of crowds by asking survey respondents to predict the voting behavior of those in their social circles \citep{galesic2018}.  The method proposed in this paper differs from all of these, and involves bringing models into direct interaction with humans in prediction markets through the actions of virtual traders. Each of these hybridization approaches is promising in its own right, and has the potential to sharpen predictive accuracy, not just for elections but for important and relatively infrequent events more generally. 



\bibliographystyle{plainnat}
\bibliography{References}

\end{document}